\documentclass[12pt]{article}

\usepackage{graphicx}

\usepackage[utf8]{inputenc}
\usepackage{amsfonts}

\usepackage{amssymb}
\usepackage{makeidx}
\usepackage{graphicx}
\usepackage{verbatim}
\usepackage{multicol}
\usepackage{multirow}
\newcommand{\la}{\lambda}

\renewcommand{\>}{{\rangle}}

\renewcommand{\]}{{\rangle\!\rangle}}
\renewcommand{\[}{{\langle\!\langle}}

\newcommand{\beq}{\begin{equation}}
\newcommand{\eeq}{\end{equation}}
\newcommand{\eq}{\end{equation}}
\newcommand{\bea}{\begin{eqnarray}}
\newcommand{\eea}{\end{eqnarray}}
\newcommand{\with}{{\quad{\rm with}\quad}}

\renewcommand{\and}{{\quad{\rm and}\quad}}

\renewcommand{\=}{\ =\ }

\newcommand{\be}{\begin{equation}}
\newcommand{\ee}{\end{equation}}

\newcommand{\bbr}{\langle\!\langle}
\newcommand{\kkt}{\kt\!\kt}

\newcommand{\pkt}{\!\!\succ\,\,}
\newcommand{\kt}{\rangle}
\begin{document}

\begin{center}

.\vspace{1cm}

 \begin{center}{\Large \bf

Quasi-Hermitian quantum mechanics
and a new class of user-friendly matrix Hamiltonians

  }\end{center}

\vspace{0.8cm}

  {\bf Olaf Lechtenfeld}$^{a}$ and
  {\bf Miloslav Znojil}$^{b,c,d}$

\end{center}

$^a$ Institut f\"ur Theoretische Physik and Riemann Center for Geometry and Physics,
Leibniz Universit\"at Hannover, Appelstrasse 2, 30167 Hannover, Germany

$^b$ The Czech Academy of Sciences,
 Nuclear Physics Institute,
 Hlavn\'{\i} 130,
250 68 \v{R}e\v{z}, Czech Republic\footnote{{e-mail:
znojil@ujf.cas.cz}}

$^c$ Department of Physics, Faculty of
Science, University of Hradec Kr\'{a}lov\'{e}, Rokitansk\'{e}ho 62,
50003 Hradec Kr\'{a}lov\'{e},
 Czech Republic
 
$^d$ Institute of System Science, Durban University of Technology,
Durban, South Africa

\section*{Abstract}

In the conventional Schr\"{o}dinger's formulation of quantum mechanics
the unitary evolution of a state $\psi$ is controlled, in Hilbert space ${\cal L}$, by a
Hamiltonian $\mathfrak{h}$
which must be self-adjoint.
In the recent, ``quasi-Hermitian'' reformulation of the theory
one replaces $\mathfrak{h}$
by its isospectral but non-Hermitian avatar $H  = \Omega^{-1}\mathfrak{h}\Omega$
with $\Omega^\dagger\Omega = \Theta \neq I$.
Although
acting in another, manifestly unphysical Hilbert space ${\cal H}$,
the amended Hamiltonian $H \neq H^\dagger$ can be perceived as self-adjoint
with respect to the amended inner-product metric $\Theta$.
In our paper motivated by a generic technical ``user-unfriendliness''
of the non-Hermiticity of $H$
we introduce and describe a specific new family of
Hamiltonians $H$ for which the metrics $\Theta$
become available in closed form.

\section{Motivation.}

\medskip
The concept of a
``user-friendly'' quantum Hamiltonian is vague: In
multiple realistic applications of
Schr\"{o}dinger
equation
 \be
 {\rm i}\frac{d}{dt}\,|\psi(t)\pkt=\mathfrak{h}\,|\psi(t)\pkt\,,
 \ \ \ \ \ |\psi(t)\pkt \in {\cal L}
 \label{prese}
 \ee
people, typically, require that the user-friendly Hamiltonians
should be just self-adjoint in Hilbert space ${\cal L}$
and, in calculations, ``sufficiently easily'' diagonalizable.

In practice,
even the latter
two elementary requirements may prove difficult
to satisfy {\em simultaneously}. One of the very old
(but still quite persuasive) examples of such a ``user-unfriendly''
self-adjoint (but difficult to diagonalize) many-body
$\mathfrak{h}=\mathfrak{h}^\dagger$
has been studied by Dyson \cite{Dyson}.
The point is that he was still able to re-classify his
problem as ``user-friendly''.

What Dyson managed to find
was a new strategy of solving
Eq.~(\ref{prese})
via a simulation of certain decisive,
diagonalization-preventing multi-fermionic
correlations.
In technical terms, the success resulted from his judicious
guess of a Hilbert-space-amending ansatz
 \be
 |\psi(t)\pkt
 =\Omega\,|\psi(t)\kt\,,\ \ \ \  |\psi(t)\kt \in {\cal H}\,.
 \label{usima}
 \ee
In contrast to the common practice,
the invertible mapping $\Omega$ was chosen {\em non-unitary},
i.e., such that  $\Omega^\dagger\Omega\neq I$.
As a consequence, the original Dyson's ``unfriendly''
Schr\"{o}dinger
equation (\ref{prese}) living in a fermionic Hilbert space ${\cal L}$
acquired an equivalent but decisively user-friendlier form
 \be
 {\rm i}\frac{d}{dt}\,|\psi(t)\kt=H\,|\psi(t)\kt\,,
 %\ \ \ \ \ |\psi(t)\kt \in {\cal H}\,,
 \ \ \ \ \ H = \Omega^{-1}\,\mathfrak{h}\,\Omega\, \neq H^\dagger\,.
 \label{amprese}
 \ee
Here,
one only had to replace the conventional Hermiticity
constraint
$\mathfrak{h}=\mathfrak{h}^\dagger$
as valid in ${\cal L}$
by the so called quasi-Hermiticity \cite{Geyer}
{\it alias\,} $\Theta-$quasi-Hermiticity \cite{ali} property
 \be
 H^\dagger\,\Theta=\Theta\,H\,,
 \ \ \ \ \
 \Theta=\Omega^\dagger\Omega\,
 \label{obeye}
 \ee
of the new, isospectral Hamiltonian operator
$H$ acting in another Hilbert space ${\cal H}$.

Incidentally (or rather on purpose),
the new Hilbert space ${\cal H}$ of the Dyson's multifermionic model
was required to be the space of
the so called ``effective'' or ``interacting'' bosons
(cf. also the
later successful use of the idea in nuclear physics \cite{Jenssen}).
This was, in some sense, the reason why the
manifestly non-Hermitian $H$ happened to become,
not quite expectedly, ``diagonalization-friendlier''.

In 1992, the ideas behind the old
Dyson's quasi-Hermitian version (\ref{amprese})
of the interacting boson model
have been reinterpreted in review \cite{Geyer}.
Its authors proposed to invert the flowchart and to
build, systematically,
the whole new family of
user-friendly quantum models by
{\em preselecting\,} a set of suitable {\em non-Hermitian\,}
observables $\Lambda_j$ which all obeyed an analogue
of Eq.~(\ref{obeye}),
 \be
 \Lambda_j^\dagger\,\Theta=\Theta\,\Lambda_j\,,
 \ \ \ \ \
 j=0,1,\ldots,K
 \label{zobeye}
 \ee
and which,
naturally, included
also the upper-case Hamiltonian as special case,
$\Lambda_0=H$.

A few years later the idea has been further developed to its perfection
and wide-spread appeal by Bender with Boettcher \cite{BB}.
Unfortunately, a weakness
of the specific Bender's and Boettcher's
ordinary differential Hamiltonians $H_{(BB)}$
appeared to be deep.
After a rigorous mathematical analysis \cite{Siegl} it
has been proved that up to a few remarkable
exceptions \cite{BG}, most of the
popular toy-model operators $H_{(BB)}$ cannot
be assigned, for certain subtle but persuasive reasons,
{\em any\,} admissible
Hilbert-space metric $\Theta_{(BB)}$.

Along the lines indicated in monograph \cite{book}
the only mathematically consequent
way out of the crisis has been found in a return to the
more restrictive version of the innovative
quasi-Hermitian quantum mechanics of review \cite{Geyer}
in which
all of the eligible non-Hermitian
operators of observables $\Lambda_j$ were required bounded in ${\cal H}$.

These developments put
in doubt
the possibility of an
internally consistent applicability of the
quasi-Hermitian quantum theory to
all of the
truly user-friendly linear differential candidates for the Hamiltonians
which happened to be unbounded.
One of the consequences was that
the attention of many quantum physicists
had to be
redirected back to the
models with finite-dimensional Hamiltonian matrices
(cf. their characteristic samples
in \cite{Geyer,zno11,zno12}).

The major part of the currently developing non-Hermitian picture of dynamics
had to be realized in finite-dimensional
Hilbert spaces.
One of the collateral practical loses was that even then,
most of the restricted matrix or bounded-operator
models admitted just a brute-force numerical analysis, the results of
which often happened to be impractical, mainly due to the
non-Hermitian  matrix nature of
the ${M}$ by ${M}$ Hamiltonians $H^{({M})}$.

Our present project of a search
for the
user-friendly quantum Hamiltonians
$H^{({M})}$
with {\em any\,} matrix
dimension ${M} \leq \infty$
was motivated precisely by the latter discouraging observations.

\section{Zig-zag matrix algebra.}

\medskip
Some encouraging preliminary results
as well as
an immediate inspiration of our present paper
can be found in
Ref. \cite{zno23} (to be cited as paper P1 in what follows).
One of us noticed there that
there exist several
parallels between the simplest ``irreducible'' special case of a non-diagonal
{\em self-adjoint\,} (i.e., Hermitian) matrix $\mathfrak{h}^{({M})}$
(i.e., between the
tridiagonal matrix defined, in general, in terms of $3{M}-2$ independent
real parameters) and its simplest non-Hermitian analogues $H^{({M})}$ having
one of the two alternative
sparse-tridiagonal forms
 \be
 H^{({M})}_{ZZ}(\vec{a},\vec{c})=
 \left[ \begin {array}{cccccc}
     {\it a_1}&0&0&0&\ldots&
 \\\noalign{\medskip}{\it c_1}&{\it a_2}&{\it c_2}&0&\ldots&
 \\\noalign{\medskip}0&0&{\it a_3}&0&0&\ldots
 \\\noalign{\medskip}0&0&{\it c_3}&{\it a_4}&{\it c_4}&\ddots
 \\\noalign{\medskip}\vdots&\vdots&0&0&{\it a_5}&\ddots
 \\\noalign{\medskip}&&\vdots&\ddots&\ddots&\ddots
 \end {array} \right]\,\ \ \ \ \ \ \ \ 
 \ee
 \be
 \ \ \ \ \ \ \ \ 
 H^{({M})}_{TZ}(\vec{a},\vec{c})=
 \left[ \begin {array}{cccccc}
     {\it a_1}&{\it c_1}&0&0&\ldots&
 \\\noalign{\medskip}0&{\it a_2}&0&0&\ldots&
 \\\noalign{\medskip}0&{\it c_2}&{\it a_3}&{\it c_3}&0&\ldots
 \\\noalign{\medskip}0&0&0&{\it a_4}&0&\ddots
 \\\noalign{\medskip}\vdots&\vdots&0&{\it c_4}&{\it a_5}&\ddots
 \\\noalign{\medskip}&&\vdots&\ddots&\ddots&\ddots
 \end {array} \right]\,
 \label{defzz}
 \ee
called ``zig-zag matrix''
and ``transposed zig-zag matrix'', respectively.

For the reasons as explained in
\cite{zno23} it makes sense to notice that the spectra of $H$s
in (\ref{defzz}) coincide with the
elements $\{a_1,a_2,\ldots,a_{M}\}$ of the vector $\vec{a}$
(cf. Lemma Nr. 5 in P1).
This enables one to
impose the spectrum-reality constraint ${\rm Re\,}a_j=a_j\neq 0$
and to
treat these matrices as
quasi-Hermitian candidates for quantum
bound-state Hamiltonians
compatible with Eq.~(\ref{obeye}).
As a consequence, the analogy with the tridiagonal Hermitian
matrices
is even closer because every matrix $H$
of Eq.~(\ref{defzz}) becomes also defined strictly
in terms of $3{M}-2$ independent
real parameters.

Among several other remarkable properties of the latter
(i.e., ZZ or, analogously, TZ) matrices let us also mention
that they form the two respective sets
which are closed with respect
to the multiplication (cf. Lemma Nr. 6 in paper P1).
For our present purposes it makes also sense, as it did in P1,
to simplify the discussion and to
work just with the real
and invertible ZZ or TZ matrices $H$
with non-degenerate spectra, i.e., just with the $2{M}-1$ nontrivial
real matrix elements such that ${\rm Im\,}c_j=0$
and $a_j \neq a_k$ at all $j$ and $k \neq j$.

\section{Quantum zig-zag-matrix Hamiltonians.}

\medskip
Besides the wealth of the remarkable algebraic
characteristics of the ZZ and/or TZ sets (cf., e.g.,
the closed-form matrix-inversion rule of
Lemma Nr. 8 in P1), the main emphasis has to be
put upon their use, in the role of
the user-friendly operators of observables,
in quantum physics. In this setting
it is necessary to emphasize
the emergence of
their two formal merits. The first one
lies,
after both of the alternative ZZ or TZ choices of the Hamiltonian,
in the closed-form solvability
of the time-independent Schr\"{o}dinger equations
(cf. Lemmas Nr. 1, 2 and 3 in P1). This means that not only the spectra
but also the states become available in closed form.

The second formal merit is even more important. In P1 this merit was
given the form of the main Theorem Nr. 4 stating that for any given ZZ or TZ $H$,
{\em all\,} of the operators $\Theta$ compatible with Eq.~(\ref{obeye})
(i.e., solving this equation and
making the matrix $H$
quasi-Hermitian)
become available in
an explicitly specified pentadiagonal-matrix form.

The latter result is of paramount importance in applications.
For two reasons.
The first one is physical, implying
the conventional
probabilistic interpretation of the theory. This enables us to call
the matrix $H$
the Hamiltonian of a quantum system.
The second, mathematical reason is almost trivial: Whenever one tries to follow the
quasi-Hermitian model-building recipe of review \cite{Geyer}
and whenever one considers
a suitable non-Hermitian ZZ or TZ candidate $H^{({M})}$
for the Hamiltonian,
one has no problem with the necessary proof
of the
reality of the spectrum
{\it alias\,} of the
observability of all of the bound-state energies
(i.e., one simply recalls that  ${\rm Im\,}a_j=0$
at all $j$).

This guarantees the
existence of at least one matrix of metric $\Theta$
with which
the quasi-Hermiticity constraint (\ref{obeye})
becomes satisfied.
In fact, the assignment of $\Theta$
to a preselected $H$ (with real spectrum)
is one of the main technical challenges encountered during
any successful
construction of a quasi-Hermitian physical quantum
system \cite{ali}. Hence, its closed-form feasibility
as mentioned above
is in fact one of the decisive merits of the ZZ and/or TZ matrix models
of paper P1.

In the context of physics, the assignment $H \to \Theta(H)$ has to be
perceived as ambiguous. In this sense one can speak about
another merit of the closed-form solvability of Eq.~(\ref{obeye}).
Indeed, its closed-form solutions as prescribed by Theorem Nr. 4 of paper P1
form an ${M}-$parametric set, i.e.,
they represent an exhaustive and entirely general
solution of the eligible-inner-product problem.
Covering all of the
existing complementary choices of the other observables $\Lambda_j$.
In this sense, relations (\ref{zobeye}) can be recalled as the additional input
information about dynamics fixing the parameters and removing completely the ambiguity
of $\Theta$ \cite{Geyer}.

It is worth adding that the essence of the availability of the ZZ or TZ
inner-product metric $\Theta$ lies in its above-mentioned
factorization $\Theta=\Omega^\dagger\Omega$. It
is related
to a slightly re-arranged formula
taken from Eq.~(\ref{amprese}),
i.e.,
to relation
 \be
 H^\dagger\,\Omega^\dagger_0 = \Omega^\dagger_0\,\mathfrak{h}_0\,.
 \label{rein}
 \ee
The subscripted zero indicates here that
a ``special'' Hermitian partner matrix $\mathfrak{h}_0$ is chosen diagonal.

According to Lemma Nr. 3 of P1 we may then start, say, from the left-hand-side
Hamiltonian
$H^{({M})}_{ZZ}(\vec{a},\vec{c})$ of Eq.~(\ref{defzz}),
and we may re-write Eq.~(\ref{rein}) in the form of equation Nr. (13) of P1,
i.e.,
in its explicit Schr\"{o}dinger-equation form
 \be
 H^{({M})}_{TZ}(\vec{a},\vec{c})\,|\psi_n\kkt =E_n\,|\psi_n\kkt \,,\ \ \ \
 E_n=\left (\mathfrak{h}_0\right )_{nn}
 \label{drein}
 \ee
of which the solutions are known in closed form,
\be
 \{
 |\psi_1^{}\kkt,|\psi_2^{}\kkt,\ldots,|\psi_M^{}\kkt
 \}
 =H_{(TZ)}(\vec{p},\vec{q})\,,
 \label{normTackocis}
 \ee
With the normalization $p_j=1$ (at all $j$) one gets
 \be
 q_k=-c_k/(a_k-a_{k+1})\,,\ \ \ \ \ k = 1,2,\ldots,{M}-1\,.
 \label{defel}
 \ee
(cf. Lemma Nr. 3 in P1).

The climax of the story then comes with the formula
{\it alias\,} definition
 \be
 \Theta=\Theta(\kappa^2_1\,,\kappa^2_2\,, \ldots,\kappa^2_M)
 =\sum_{n=1}^M\,|\psi_n^{}\kkt\,\kappa^2_n\,\bbr \psi_n^{}|\,.
 \label{odkap}
 \ee
in which all of the real parameters $\kappa^2_j\neq 0$ are arbitrary
(cf. \cite{SIGMAdva}).

It is possible to conclude that once we finally
have the correct physical operator of the inner-product metric (\ref{odkap})
at our disposal,
we are prepared to build the consistent quantum-mechanical models
along the more or less conventional lines and, moreover, in a strictly non-numerical manner.
In these models the ``obligatory'' requirement of the self-adjointness
of all of the eligible and relevant observables $\Lambda_j$ becomes
merely generalized and replaced
by
the $\Theta-$quasi-Hermiticity rule (\ref{zobeye}).

\section{Generalized zig-zag matrix algebras.}

\medskip
In the light of the above-outlined results we came to the conclusion that it
might make sense to search for the generalizations of the zig-zag
matrices. One of the results of this search is to be presented in
what follows.
We will show that and how the numerous
user-friendliness
features of the zig-zag models  can be generalized.

For this purpose\footnote{The discussion is easily extended
to odd dimensions.}
let us consider a Hilbert space of finite even dimension $2m$,
i.e.~$\mathbb{C}^{2m}$. We define a solvable class of
non-hermitian model Hamiltonians by
\begin{equation}
H \= \Lambda\ +\ N \quad\with \Lambda\ \textrm{diagonal} \and N\
\textrm{nilpotent of degree 2}\ .
\end{equation}
Furthermore, we impose the simplification that $N$ is built entirely from nilpotent
$2{\times}2$ blocks $B$ such that $B_{11}=B_{21}=B_{22}=0$.
%
%$\bigl(\begin{array}{cc} 0 & * \\ 0 & 0 \end{array}\bigr)$
The matrix $N$ itself need not have to be of an upper or lower block triangular form.
Still, any two such matrices multiply to zero. To bring out this $2{\times}2$ structure,
we label the row and column entries as follows,
\begin{equation}
\bigl\{ +1, -1, +2, -2, +3, -3, \ldots, +i, -i, \ldots, +m, -m \bigr\}\ ,
\end{equation}
so that the index~$i=1,\ldots,m$ numerates the blocks. With this labelling, the matrix entries are
\begin{equation}
\Lambda \= \textrm{diag}\bigl(
\la_{+1},\la_{-1},\la_{+2},\la_{-2},\ldots,\la_{+m},\la_{-m} \bigr)
\ee
and
 \be
N_{+i+j}=N_{-i-j}=N_{-i+j}=0\ ,\quad N_{+i-j}=:n_{ij}\ .
\end{equation}
Abbreviating the matrix entries by $\vec{\la}$ and $\hat{n}$ respectively, we may characterize
such matrices as $H(\vec{\la},\hat{n})$. Explicitly, they take the form
\begin{equation}
H(\vec{\la},\hat{n}) \= \left (\begin{array}{ccccccccc}
\la_{+1} & n_{11} & 0 & n_{12} & 0 & n_{13} & \cdots & 0 & n_{1m} \\
0 & \la_{-1} & 0 & 0 & 0 & 0 & \cdots & 0 & 0 \\
0 & n_{21} & \la_{+2} & n_{22} & 0 & n_{23} & \cdots & 0 & n_{2m} \\
0 & 0 & 0 & \la_{-2} & 0 & 0 & \cdots & 0 & 0 \\
0 & n_{31} & 0 & n_{32} & \la_{+3} & n_{33} & \cdots & 0 & n_{3m} \\
0 & 0 & 0 & 0 & 0 & \la_{-3} & \cdots & 0 & 0 \\
\vdots & \vdots & \vdots & \vdots & \vdots & \vdots & \ddots & \vdots & \vdots \\
0 & n_{m1} & 0 & n_{m2} & 0 & n_{m3} & \cdots & \la_{+m} & n_{mm} \\
0 & 0 & 0 & 0 & 0 & 0 & \cdots & 0 & \la_{-m}
\end{array} \right )\ .
\end{equation}
The odd-dimensional case can be treated in the same fashion via enlarging the matrix~$H$ by
a zero row and zero column. In this way, it is a special even-dimensional system with
$\la_{m-}=n_{im}=0$.

This class of matrices has interesting properties.
Firstly, it is closed under multiplication, since left or right multiplication with a diagonal
matrix conserves the structure of~$N$,
\begin{equation}
(\Lambda+N)\,(\Lambda'+N') \= (\Lambda\Lambda') +
(\Lambda\,N'+N\Lambda') \= \Lambda'' + N''\ .
\end{equation}
%\begin{equation}
%H(\vec{\la},\hat{n})\,H(\vec{\la}',\hat{n}') \= H(\vec{\la}'',\hat{n}'') \quad\with
%\la_{\pm i\pm j}'' = \la_{\pm i\pm j} \la_{\pm i\pm j}' \and
%n_{ij}'' = \la_{+i}n_{ij}' + n_{ij}\la_{-j}'\ .
%\end{equation}
Even stronger, any pattern of vanishing $2{\times}2$ blocks, i.e.~some collection of
vanishing~$n_{ij}$, is preserved under matrix multiplication.
If none of the eigenvalues $\la_{\pm i}$ is vanishing, an inverse can be constructed,
\begin{equation}
(\Lambda + N)^{-1} \= \Lambda^{-1} - \Lambda^{-1} N\,\Lambda^{-1}\ .
\end{equation}

\section{Generalized zig-zag-matrix Hamiltonians.}

\medskip
Once we move to the applications of the generalized zig-zag-matrices
in quantum mechanics,
the eigensystem is also found easily. Obviously, the eigenvalues are
simply the diagonal entries~$\la_{\pm i}$ of~$\Lambda$. Let us
arrange the eigenket columns $|{\pm}i\>$ defined (up to
normalization) by
\begin{equation} \label{schr}
H\,|{\pm}i\> \= \la_{\pm i}\,|{\pm}i\>
\end{equation}
into a $2m{\times}2m$ matrix
\begin{equation}
Q \= \bigl(\ |{+}1\> \ |{-}1\> \ |{+}2\> \ \ldots \ |{-}m\> \ \bigr) \= {\bf 1} + \bar{N}\ ,
\end{equation}
where the second equation defines an ansatz, which gives $Q$ the same form as~$H$.
The choice of the unit matrix as the diagonal part fixes a normalization of the eigenkets.
The defining equation~\ref{schr} for~$Q$ is easily solved,
\begin{equation}
H\,Q = Q\,\Lambda \qquad\Leftrightarrow\qquad [ \bar{N},\Lambda ] =
N \qquad\Leftrightarrow\qquad \bar{N}_{+i-j} =
-n_{ij}/(\la_{+i}{-}\la_{-j})\ .
\end{equation}
Explicitly, the matrix of eigencolumns reads
\begin{equation}
Q = \left (\begin{array}{ccccccccc}
1 & q_{+1-1} & 0 & q_{+1-2} & 0 & q_{+1-3} & \cdots & 0 & q_{+1-m} \\
0 & 1 & 0 & 0 & 0 & 0 & \cdots & 0 & 0 \\
0 & q_{+2-1} & 1 & q_{+2-2} & 0 & q_{+2-3} & \cdots & 0 & q_{+2-m} \\
0 & 0 & 0 & 1 & 0 & 0 & \cdots & 0 & 0 \\
0 & q_{+3-1} & 0 & q_{+3-2} & 1 & q_{+3-3} & \cdots & 0 & q_{+3-m} \\
0 & 0 & 0 & 0 & 0 & 1 & \cdots & 0 & 0 \\
\vdots & \vdots & \vdots & \vdots & \vdots & \vdots & \ddots & \vdots & \vdots \\
0 & q_{+m-1} & 0 & q_{+m-2} & 0 & q_{+m-3} & \cdots & 1 & q_{+m-m} \\
0 & 0 & 0 & 0 & 0 & 0 & \cdots & 0 & 1
\end{array}\right )
\end{equation}
with $$ q_{+i-j} = -\frac{n_{ij}}{\la_{+i}{-}\la_{-j}}\,.$$
This solution is only viable if $\la_{+i}\neq\la_{-j}$ whenever $n_{ij}$ is
nonzero\footnote{In such a non-admissible situation our matrix features at least one Jordan block,
rendering it non-diagonalizable.}.
In general though, eigenvalue degeneracies are allowed.

The special case of $n_{ij}=0$ except for $j{=}i$ and $j{=}i{+}1$
(i.e.~only secondary and quaternary sparse upper diagonals) reduces
to the zig-zag matrices discussed in~\cite{zno23}. They are
reproduced by a basis permutation $+i\leftrightarrow-i$, which
merely swops the labels in each $2{\times}2$ block. The transpose
matrix~$H^\top=\Lambda+N^\top$ is obtained by transposing the
$2{\times}2$ blocks internally and also blockwise in~$H$, hence
$N^\top_{-j+i}=n_{ij}$ and others vanish. 

For
 $$
 \widetilde{Q} = \bigl( \
|{+}1\] \ \ldots \ |{-}m\] \ \bigr) \= {\bf 1} + \widetilde{\bar{N}}
 $$
the eigenvalue problem
\begin{equation}
H^\top |{\pm}i\] \= \la_{\pm i}\,|{\pm}i\]
\qquad\Leftrightarrow\qquad H^\top \widetilde{Q} =
\widetilde{Q}\,\Lambda 
\end{equation}
with the analog ansatz $\widetilde{Q}={\bf 1}+\widetilde{\bar{N}}$ has the solution
\begin{equation}
\widetilde{\bar{N}} \= -\bar{N}^\top \qquad\Leftrightarrow\qquad
\widetilde{\bar{N}}_{-j+i} \= n_{ij}/(\la_{+i}{-}\la_{-j})\ .
\end{equation}

The quasi-Hermiticity of~$H$ defines a physical inner-product metric~$\Theta$ in our Hilbert space, via
\begin{equation}
H^\top \Theta \= \Theta\,H\ .
\end{equation}
In the same manner as above, all of the admissible metrics are given by formula
\begin{equation} \label{Theta1}
\Theta\bigl(\{\kappa_{\pm i}^2\}\bigr) \= \sum_{i=1}^m
\Bigl(\, |{+}i\] \, \kappa_{+i}^2 \, \[ {+}i|\ +\ |{-}i\] \, \kappa_{-i}^2 \, \[ {-}i| \,\Bigr)\,
\end{equation}
parametrized by $2m$ nonnegative parameters~$\kappa_{\pm i}$.
Collecting these parameters in a diagonal matrix
\begin{equation}
K^2 \= \textrm{diag} \bigl( \kappa_{+1}^2, \kappa_{-1}^2, \kappa_{+2}^2, \ldots, \kappa_{-m}^2 \bigr) \ ,
\end{equation}
the metric matrix can be written as
\begin{equation}
\Theta \= \widetilde{Q}\,K^2 \widetilde{Q}^\top \= K - \bar{N}^\top K^2 - K^2\bar{N} + \bar{N}^\top\bar{N}
\end{equation}
where $\bar{N}$ is the off-diagonal part of the solution~$Q$.

As long as $\bar{N}$ and $\bar{N}^\top$ belong to two disjoint classes of nilpotent matrices, this formula
in general produces a full $2m{\times}2m$ matrix. For the class of zig-zag matrices, which are empty
beyond the quaternary diagonal, in our basis the metric becomes block tridiagonal, i.e.~it is heptadiagonal,
but both secondary, tertiary and quaternary diagonals remain sparse. Passing to the `zig-zag basis'
the quaternary diagonals can be removed for the prize of non-sparse secondary ones.

%\newpage
%%%

\end{document}